\documentclass[aps,amsfonts,nofootinbib,article,twocolumn]{revtex4}
\usepackage{amsmath}
\usepackage{amsfonts}
\usepackage{amssymb}
\usepackage{pbsi}
\usepackage[T1]{fontenc}
\usepackage{hyperref}

\def\openone{\leavevmode\hbox{\small1\kern-3.8pt\normalsize1}}
\def\N{\leavevmode\hbox{ Z \kern-8 pt\normalsize{Z}}}
\def\openone{\leavevmode\hbox{\small1\kern-3.8pt\normalsize1}}
\def\openJ{\leavevmode\hbox{J \kern-9.5pt\normalsize J}}
\def\openS{\leavevmode\hbox{ S \kern-9.3pt\normalsize S}}
\newcommand{\bb}{\begin{equation}}
\newcommand{\ee}{\end{equation}}
\newcommand{\eqb}{\begin{eqnarray}}
\newcommand{\eqf}{\end{eqnarray}}

\usepackage{color}

\begin{document}

\title{Geometrical Unification of Gravitation and Electromagnetism}

\author{Sergio A. Hojman}
\email{sergio.hojman@uai.cl}
\affiliation{Departamento de Ciencias, Facultad de Artes Liberales,
Universidad Adolfo Ib\'a\~nez, Santiago 7491169, Chile.}
\affiliation{Centro de Investigaci\'on en Matem\'aticas, A.C., Unidad M\'erida, Yuc. 97302, M\'exico}
\affiliation{Departamento de F\'{\i}sica, Facultad de Ciencias, Universidad de Chile,
Santiago 7800003, Chile.}
\affiliation{Centro de Recursos Educativos Avanzados,
CREA, Santiago 7500018, Chile.}

\begin{abstract}
A theory which unifies gravitation and electromagnetism (GUGE) is presented. This new theory is based on a recent redefinition of proper time. The 5--dimensional metric which arises is similar but not equivalent to the Kaluza--Klein (KK) one. Differences follow. The GUGE metric is deduced while the KK metric is postulated. In the GUGE field theory there is no need to impose either the ``cylinder'' or ``curling of coordinates'' conditions, because they are direct consequences of the GUGE formalism. The GUGE field equations are fully equivalent to Einstein--Maxwell equations, while KK field equations are not. The GUGE 5--dimensional (geodesic) equations are equivalent to the 4--dimensional (non--geodesic) equations for a charged particle moving in the presence of gravitational and electromagnetic fields, unlike the KK 5--dimensional (geodesic) equations which are not. No extra scalar field appears in GUGE. The physical interpretation of the fifth dimension and of the role of the extra field in KK (internal coordinate in GUGE) are totally different in both approaches. Finally, GUGE results include electric charge conservation, electric charge quantization and electric charge contribution to the energy of charged particles even in the absence of electromagnetic fields, which implies (the observed fact) that there are no massless electrically charged particles in Nature, unlike the prevailing treatments of KK theories.
\end{abstract}


\maketitle
\section{Introduction}

Several successful examples of unification theories may be listed, starting from Newton's unified explanation of gravitational phenomena both at human and astronomical scales, Maxwell's unification of electric and magnetic phenomena, Glashow, Salam and Weinberg electro--weak unification and the Standard Model of elementary particles to name a few extremely well known ones.

In the case of the unification of gravitation and electromagnetism two review articles written by Goenner \cite{goenner1, goenner2} give a clear perspective of both the relevance of the subject and of the efforts and attempts made for over a century to construct such a theory.

The geometrical unification of gravitation and electromagnetism (GUGE) theory presented here starts from an idea related to the description of the motion of a test particle subject to gravitational, electromagnetic and Yang--Mills interactions. The idea is inspired in a recent proposal of proper time redefinition \cite{propertime} which follows the path of a unified field theory described in terms of a generalized Einsten--Hilbert Lagrangian based on the construction of a Finsler metric \cite{rund}. The metric so constructed turns out to be Riemannian, which is, of course a special case of a Finsler metric. The new definition of proper time gives naturally rise to the construction of a Riemann metric, which introduces all known fundamental interactions in its definition in contradistinction to the usual one which incorporates gravitational interactions only.

Metrics are usually defined in geometrical terms using the concept of ``length'', as it is done in the Riemannian case. In the case in which the particle interacts with gravitational interactions only, it turns out that the ``length'' is exactly proportional to the action of the Lagrangian which describes the dynamics of a massive neutral point particle moving on a gravitational background. The ``length'' must be a homogeneous function of degree one in the coordinate differentials, which in turn means that the Lagrangian must be a homogeneous function of degree one on the velocities, in order to be able to define a reasonable metric. It turns out that the gravitational part of the Lagrangian is a homogeneous function of degree one on the velocities. Moreover, the addition to the Lagrangian of new terms which describe both electromagnetic and general Yang--Mills interactions satisfies that condition too and, in fact, these new interaction terms are linear in the velocities \cite{propertime,bal1,bal2,casa1,casa2}. Nevertheless, the usual action which describes the motion of a massive, spinless, charged test particle in the presence of an external electromagnetic field {\it{is not gauge invariant}} and, therefore, cannot be used to define proper time. The introduction of internal particle coordinates associated to electric and Yang--Mills charges allows for the definition of a new action which gives rise to electric charge and isotopic spin magnitude conservation, in addition to the same equations of motion for the charged particle. This new action is used to redefine proper time \cite{propertime}.

In order to incorporate non--gravitational interactions, additional complex coordinates (related to electromagnetic and to Yang--Mills charges) which supplement the spacetime ones are included. The Lagrangian then, is a real and {\it {electromagnetic gauge invariant}} function which depends both upon real (spacetime coordinates) and complex (electromagnetic and Yang--Mills) charges variables.

For a Lagrangian which is a homogeneous function of degree one in the velocities one may define a metric as one half of the second derivatives with respect to the velocities of the square of the Lagrangian which gives rise to the equations of motion of a relativistic point particle interacting with all of the fundamental interactions \cite{propertime}. This metric coincides with the Riemann metric of spacetime, if the only interactions considered are the gravitational ones.

In this work, a simplification of the full Lagrangian for gravitation and electromagnetism is achieved by using two real variables instead of the two complex ones presented in \cite{propertime}. For the electromagnetic case, the resulting five--dimensional Riemmannian metric, turns out to be direction independent and {\it{looks}} exactly equal to the usual (5 dimensional) Kaluza--Klein (KK) metric.\\

The differences between the KK and GUGE approaches are summarized in a Table at the end of the article.\\

\section{Geometrical Unification of Gravitation and Electromagnetism}

Proper time redefinition \cite{propertime} starts from the observation that the massive spinless point particle action $S$ and proper time $\tau$ are proportional, as it is well known, where
\begin{equation}
L^R=m_0\ c\ ({g^R}_{\mu \nu} (x^\alpha)\ u^\mu u^\nu)^{1/2}\, = m_0\ c\ \frac{d s}{d \lambda}, \label{lagpoint}
\end{equation}
where $m_0$ is the particle's rest mass, ${g^R}_{\mu \nu} (x^\alpha)$ the Riemannian spacetime metric and the action is
\begin{eqnarray}
S^R & & = \int_{i}^{f} L d \lambda = m_0\ c\ \int_{i}^{f}  \frac{d s}{d \lambda} d \lambda =  \nonumber \\  & & = m_0\ c\  \int_{i}^{f}  ds = m_0\ c^2\  \int_{i}^{f} d\tau, \label{actionpoint}
\end{eqnarray}
and $\lambda$ is an arbitrary parameter.
Set $m_0=1$ and $c=1$ and define, as usual, the canonical momentum ${P^R}_\mu$
\begin{equation}
{P^R}_\mu \equiv \frac{\partial L^R}{\partial u^\mu}= \frac{g_{\mu \nu} u^\nu}{L^R}= g_{\mu \nu} \frac{d x^\nu}{ds} , \label{momentum}
\end{equation}
which is $\lambda$--reparametrization invariant and homogeneous of degree zero on the velocities.
Note that
\begin{equation}
{P^R}_\mu {P^R}^\mu \equiv - 1.\label{momentumconstraint}
\end{equation}
The usual Lagrangian $L_{EM}$ for a charged massive spinless point particle interacting with an external electromagnetic field described in terms of vector potential $A_\rho (x^\beta)$ is
\begin{equation}
L_{EM}=({g^R}_{\mu \nu} \ u^\mu u^\nu)^{1/2}\ + q u^\rho A_\rho =\frac{ds}{d\lambda}+ q u^\rho A_\rho , \label{lagcharged}
\end{equation}
which yields the very well known equations of motion
\begin{equation}
\frac{D ({g^R}_{\mu \nu}\frac{d x^\nu}{ds})}{D \lambda} = q F_{\mu \nu} \frac{d x^{\nu}}{d \lambda}\ , \label{eqmchg}
\end{equation}
where $D/d \lambda$ denotes covariant derivative with respect to $\lambda$.\\

Nevertheless, the action associated to Lagrangian \eqref{lagcharged} {\it{is not}} electromagnetic gauge invariant and, consequently, cannot be thought of as proportional to proper time. This difficulty was solved in \cite{propertime} by replacing Lagrangian \eqref{lagcharged},
\begin{eqnarray}
\bar L_\theta =& & \dot s\ + \frac{i}{2}({\theta_1}^* \dot \theta_1  - \theta_1 {\dot {\theta_1}}^* + {\theta_2}^* \dot \theta_2  - \theta_2 {\dot {\theta_2}}^*) {\nonumber} \\  & & + g_e (\theta_1 {\theta_1}^*-\theta_2 {\theta_2}^*) {\dot x}^\rho A_\rho (x^\beta). \label{lagdyncharge}
\end{eqnarray}
where the dot over a variable denotes (ordinary) derivative with respect to $\lambda$. The use of complex (internal) electric charge coordinates $\theta$ and $\theta^*$ (or $Q$ and $\phi$, in what follows) for charged particles was first introduced in four articles published by Balachandran {\it{et al}} and Casalbuoni \cite{bal1,bal2,casa1,casa2}. They are the particle description equivalent of the complex functions description of  charged fields.\\

The electric charge $\bar Q \equiv {\theta_1}^*  \theta_1  - \theta_2 {{\theta_2}}^*$ is conserved for the dynamics defined by Lagrangian \eqref{lagdyncharge} as it can be easily seen by varying it with respect to $\theta_1$, ${\theta_1}^*$, $\theta_2$, and ${\theta_2}^*$. Note that the electric charge $\bar Q$ is, in principle, a function of time $t$, which is conserved by virtue of the dynamical equations of motion \cite{propertime}.

In \cite{propertime} two complex charge variables $\theta_1$ and $\theta_2$ (and their complex conjugates) were included in the Lagrangian to make sure that the conserved electric charge $\bar Q$ might assume either positive or negative (or zero) values. It is a direct matter to show that the equations of motion derived from \eqref{lagdyncharge} imply the correct equations \eqref{eqmchg} in addition to the conservation of electric charge $\bar Q$, (instead of using an externally prescribed parameter $q$).

If one proceeds to compute ${\bar L_\theta}^2$ and then define the metric by using ${g^F}_{ab}= (1/2) \partial^2 {\bar L_\theta}^2/\partial \dot{x^a} \partial \dot{x^b}$, where $a, b = 0, 1, 2, 3, 4, 5$, with $x^a = x^a$ for $a = 0, 1, 2, 3$,  $x^4 =\phi$ and $x^5 =Q$ one gets a singular 6--dimensional Finsler metric of rank 4. This seems to be a dead end as far as the goal is constructing a unified field theory {\it \`a la} Einstein \cite{goenner1, goenner2}.

In this article I take an alternative road. Define a Lagrangian based on a unique complex charge variable $\theta$ (and its complex conjugate)

\begin{eqnarray}
 L_\theta =& & \dot s\ + \frac{i}{2}({\theta}^* \dot \theta - \theta\ {\dot {\theta}}^*) + {\nonumber} \\  & & g_e \theta\ {\theta}^* {\dot x}^\rho A_\rho (x^\beta). \label{lagdyncharge2}
\end{eqnarray}

It is a straightforward matter to realize that the (positive) electric charge $Q = \theta\ {\theta}^*$ is conserved by virtue of the equations of motion for $\theta$ and $\theta^*$ derived from \eqref{lagdyncharge2}. Note that the complex coordinates $\theta$ and $\theta^*$ may be rewritten as

\begin{equation}
\theta = \sqrt{Q} e^{i \phi},\ \ \ \ \ \theta^* = \sqrt{Q} e^{-i \phi}, \label{theta}
\end{equation}

\noindent and the Lagrangian $ L_\theta$ written in terms of the new internal electric charge coordinates $Q$ and $\phi$ is

\begin{eqnarray}
L_\theta\ =\ \ \dot s\ + \ Q \left(g_e A_\mu u^\mu+ \dot {\phi}\right) . \label{lagdyncharge3}
\end{eqnarray}

The conserved charge $Q$ is positive and the equations of motion need to incorporate the fact that charged particles may have either positive or negative electric charge. One way to recover the contents of equations \eqref{eqmchg} is to write two separate equations of motion for positive and negative charges, i.e.,

\begin{equation}
\frac{D ({g^R}_{\mu \nu}\frac{d x^\nu}{ds})}{D \lambda} = Q F_{\mu \nu} \frac{d x^{\nu}}{d \lambda}\ , \label{eqmchgpos}
\end{equation}

and

\begin{equation}
\frac{D ({g^R}_{\mu \nu}\frac{d x^\nu}{ds})}{D \lambda} = - Q F_{\mu \nu} \frac{d x^{\nu}}{d \lambda}\ . \label{eqmchgneg}
\end{equation}

The electric charge conservation equation $\dot Q = 0$, and the particles equations of motion \eqref{eqmchgpos} and \eqref{eqmchgneg} may be derived by using the (matrix) Lagrangian ${\mathbb L} (x^\alpha, u^\beta, Q, \phi, \dot \phi)$

\begin{eqnarray}
{\mathbb L} (x^\alpha, u^\beta, Q, \phi, \dot \phi)\ =\ \sigma_0 \ \dot s\ +  \sigma_3 \ Q \left(g_e A_\mu u^\mu+ \dot {\phi}\right) \label{lagdynchargematrix}
\end{eqnarray}

\noindent which are written in terms of the Pauli matrices, by varying with respect to $\phi$ and $x^\alpha$.

The Pauli matrices are

\begin{center}
\noindent\(\sigma _1=\left(
\begin{array}{cc}
 0 & 1  \\
1 & 0  \\
\end{array}
\right)\),
\end{center}

\begin{center}
\noindent\(\sigma _2=\left(
\begin{array}{cc}
 0 & -i  \\
i & 0  \\
\end{array}
\right)\),
\end{center}

and

\begin{center}
\noindent\(\sigma _3=\left(
\begin{array}{cc}
 1 & 0  \\
0 & -1  \\
\end{array}
\right)\).
\end{center}

The identity matrix $\sigma_0$ is

\begin{center}
\noindent\(\sigma _0=\left(
\begin{array}{cc}
 1 & 0  \\
0 & 1  \\
\end{array}
\right)\).
\end{center}

The dynamical variables in Lagrangians \eqref{lagdynchargematrix} are $x^\mu$ and $\phi$, while $Q$ plays the role of a Lagrange multiplier.\\

Note that the electric charge $Q$ is, in principle a function of time $t$, which is conserved by virtue of the equation of motion obtained by varying $\phi$. The $x^\mu$ equations are the same as \eqref{eqmchg} (or equations \eqref{eqmchgpos} and \eqref{eqmchgneg}) except that the external parameter $q$ is replaced by the dynamically conserved electric charge $Q$.

The constraint associated to the variation of $Q$ defines $\phi$ as a path--dependent phase much in the same sense as the ones introduced by both Mandelstam  and Bia\l{}ynicki--Birula \cite{mandelstam,bb},

\begin{equation}\label{mandel}
g_e A_\mu u^\mu+ \dot {\phi}=0,
\end{equation}

or

\begin{equation}
\phi  = k  -  \int g_e A_\mu dx^\mu \label{wilson}
\end{equation}
which shows how the coordinate $\phi$ is related to ``Wilson lines'' and ``Wilson loops'' \cite{wilson}. Here $k$ is an integration constant. Note that no fields depend on $\phi$ in this approach. The particle's Lagrangian depends only on $\phi$'s velocity, $\dot{\phi}$.\\

The usual gauge transformations for the electromagnetic potentials $A_\mu(x^\alpha)$ and the phase $\phi$
\begin{eqnarray}
A'_\mu (x^\alpha)u^\mu \equiv A_\mu (x^\alpha) u^\mu+ \frac{1}{g_e} \dot \Lambda \label{gtA}
\end{eqnarray}
and
\begin{eqnarray}
\dot {\phi}' = \dot \phi(x^\alpha) - \dot \Lambda \label{gtfi}
\end{eqnarray}
for an arbitrary function $\Lambda$, leave Lagrangians \eqref{lagdynchargematrix} invariant.
Thus, these new ({\it{electromagnetic gauge invariant}}) Lagrangians (and associated actions) reproduce the usual dynamical equations for the coordinates $x^\mu$ and, additionally, yield {\it{electric charge conservation}}.\\

Define the Lagrangian $L$, by

\begin{equation}\label{L2}
L^2\equiv  \frac{1}{2}\ {\text{trace}}({\mathbb L}^2),
\end{equation}

and get

\begin{equation}\label{L21}
 L^2 = {g^R}_{\mu \nu} u^\mu u^\nu + Q^2 \left(g_e A_\mu u^\mu+ \dot {\phi}\right)^2 \ .
\end{equation}

Written in detail $L^2$ reads

\begin{equation}\label{L21}
 L^2 = {g^R}_{\mu \nu} u^\mu u^\nu + Q^2 \left({g_e}^2 A_\mu A_\nu u^\mu u^\nu+ {\dot {\phi}}^2 + 2 \dot {\phi} g_e A_\mu u^\mu \right) \ . \\
\end{equation}
Lagrangians \eqref{lagdynchargematrix} are homogeneous functions of degree one in the velocities $u^\mu$ and $\dot {\phi}$ and, therefore a five--dimensional  metric ${g^F}_{ab}$ may be defined by \cite{rund}
\begin{eqnarray}
{g^F}_{ab}(x^c,Q) \equiv \frac{1}{2} \frac{\partial^2 L^2}{\partial u^a \partial u^b} \label{gF}
\end{eqnarray}
where $a, b, c ... = 0, 1, 2, 3, 4$, because no $\dot Q$ appears in the Lagrangians \eqref{lagdynchargematrix} so that ${g^F}_{aQ}={g^F}_{Qa}=0={g^F}_{QQ}$. The first four velocities are the spacetime ones and $u^4=\dot \phi$. It is immediate to note that the metric ${g^F}_{ab}(x^c,Q)={g^F}_{ab}(x^\mu,Q)$, i.e., the Riemann metric does not depend on $\phi$. Therefore, the (regular minor of the) Riemann metric happens to be five--dimensional and $Q$ plays no dynamical role from now on.

The 5--dimensional regular Riemann metric, denoted by ${g^F}_{ab}$, is such that ${g^F}_{ab}={g^{GUGE}}_{ab}\equiv{g^{G}}_{ab}$ happens to look exactly like the Riemannian, isotropic, i.e., velocity or direction independent,  Kaluza--Klein metric with

\begin{equation}
{g^F}_{ab} = {g^{G}}_{ab}\label{gF1}
\end{equation}
where
\begin{equation}
{g^{G}}_{\mu \nu} = {g^R}_{\mu \nu}+ Q^2 {g_e}^2 A_\mu A_\nu ,\label{gKK1}
\end{equation}
\begin{equation}
{g^{G}}_{\mu 4} = {g^{G}}_{4 \mu}  = Q^2 {g_e} A_\mu ,\label{gKK2}
\end{equation}
\begin{equation}
{g^{G}}_{4 4} = Q^2 ,\label{gKK3}
\end{equation}

Note, nevertheless, that the fifth--dimension is associated to the internal electric charge phase $\phi$ while $Q$ {\it {is not a new independent field}} as it is usually assumed in the prevailing treatments of Kaluza--Klein theory \cite{goenner1, goenner2}. As it has been seen,  $Q$ is the sixth (non--dynamical) internal electric charge coordinate, which is, of course, independent of all other five coordinates $x^\mu$ and $\phi$. Note that ${g^{G}}_{ab}$ is $\phi$ independent, i.e., the cylindrical condition is identically met in this formalism.

\section{5--dimensional GUGE action}

Consider now, the action for the 5--dimensional GUGE theory. It is important to realize that the dynamical variables for the action are $x^\mu$ and $\phi$, which are the variables whose time derivatives $u^\mu$ and $\dot \phi$ appear in Lagrangians $\mathbb L_\pm$ given in \eqref{lagdynchargematrix}. The 5--dimensional metric ${g^F}_{a b}$ which is computed by taking derivatives of $(1/2) L^2$  with respect to those very velocities, does not depend on $\phi$, therefore it is not necessary to invoke the so called ``cylinder condition''. Moreover, $\phi$ is a phase so its integral over spacetime is just equal to $2 \pi$. The 5--dimensional metric ${g^F}_{a b}$ turns out to be Riemannian.

The 6 coordinates $x^\mu$, $\phi$ and $Q$ are independent and the action constructed from $L$ gets varied with respect to the first five of them only, so no derivatives of $Q$ appear in the variation of the action (because $Q$ is the sixth coordinate and it is, therefore, independent of the other five coordinates $x^\mu$ and $\phi$).
The 5--dimensional GUGE action $S^{G}$ (which is formally identical to the Kaluza--Klein one presented in \cite{ce}  and in \cite{williams}) is, using the notation of this work,
\begin{equation}
S^{G}= \int (-\det g^{G})^{1/2} R^{G} d^4 x\ d \phi, \label{KKaction}
\end{equation}
\begin{equation}
S^{G}=\ 2 \pi \int (-\det g^{G})^{1/2} R^{G} d^4 x, \label{KKaction2}
\end{equation}
because there is no $\phi$ dependence. The action is

\begin{equation}
S^{G}=\ 2 \pi \int (-\det g^{R})^{1/2} \left( Q\ R^{R}-\frac{1}{4} Q^3\ F^{\alpha \beta} F_{\alpha \beta} \right) d^4 x, \label{KKaction3}
\end{equation}

and may be rewritten as

\begin{equation}
S^{G}=\ 2 \pi  Q^3 \int (-\det g^{R})^{1/2} \left(R^{R}-\frac{1}{4}F^{\alpha \beta} F_{\alpha \beta} \right) d^4 x, \label{KKaction4}
\end{equation}

\noindent which is exactly proportional to the Einstein--Maxwell action, after a four dimensional conformal transformation ${g^R}_{\mu \nu} = Q^2 {g'^R}_{\mu \nu}$ is performed (and the $'$ is dropped). Note that $Q$ does not depend on either $x^\mu$ or $\phi$ as mentioned earlier. This approach, based on previous work \cite{propertime} allows one to deduce exactly the Einstein--Maxwell theory, unlike the Kaluza--Klein theory which produces equations which are not equivalent to the Einstein--Maxwell ones.

Note that no extra assumptions such as ``cylinder condition'' or ``curling of coordinates'' (``compactification'') are needed as the action does not depend on $\phi$ and $\phi$ is naturally curled, being a phase.

\section{5--dimensional GUGE geodesics}

Consider the GUGE Lagrangian $L^{GUGE}$ which defines the dynamics of a particle as the geodesics of the 5--dimensional GUGE metric given in \eqref{gKK1}, \eqref{gKK2}, and \eqref{gKK3}, after performing a four dimensional conformal transformation ${g^R}_{\mu \nu} = Q^2 {g'^R}_{\mu \nu}$ , dropping the $'$ and dismissing the non--dynamical factor $Q^2$,

\begin{eqnarray}
L^{GUGE} & = &  (({g^R}_{\alpha \beta}+ {g_e}^2 A_\alpha A_\beta) u^\alpha u^\beta \nonumber \\ & + &2 {g_e} A_\alpha u^\alpha \dot \phi + {\dot {\phi}}^2)^{1/2}\equiv \frac{d\sigma}{d\lambda}
\end{eqnarray}
which also defines the new length differential $d\sigma$ which, in turn, is proportional to a newly defined (5--dimensional) proper time differential \cite{propertime}.

The Lagrangian $L^{GUGE}$ does not depend on $\phi$, therefore its canonically conjugated momentum $P_\phi$ is equal to a constant of motion $K$,

\begin{equation}\label{mande1}
P_\phi \equiv \frac{\partial L^{GUGE}}{\partial \dot \phi}=({g_e} A_\alpha \frac{d x^\alpha}{d \sigma}+\frac{d \phi}{d \sigma})=K.
\end{equation}
Note that the derivatives in \eqref{mande1} are taken with respect to the new proper time $\sigma$.

Compute the momentum $P_\mu$ which appears when varying $L^{GUGE}$ with respect to $x^\mu$,

\begin{equation}\label{gugeeq2}
P_\mu ={g^G}_{\mu a} \frac{d x^a}{d\sigma},
\end{equation}
or
\begin{equation}\label{gugeeq2}
P_\mu ={g^R}_{\mu \beta} \frac{d x^\beta}{d\sigma}+ A_\mu g_e (g_e A_\beta \frac{d x^\beta}{d\sigma}+ \frac{d \phi}{d\sigma}).
\end{equation}

Finally, use Eq.\eqref{mande1} to get

\begin{equation}\label{gugeeq3}
P_\mu ={g^R}_{\mu \beta} \frac{d x^\beta}{d\sigma}+ g_e K A_\mu ,
\end{equation}
which is exactly the 4--dimensional expression for the momentum as computed by using $L_\theta$ given by \eqref{lagdyncharge3} except that the derivative is taken with respect the new proper time $\sigma$ and the conserved charge is now $K$ (instead of $q$ or $Q$). Note that $K$ may be either positive or negative.

Similarly, one gets

\begin{equation}
\frac{\partial L^{GUGE}}{\partial x^\mu}= \frac{1}{2} {g^R}_{\alpha \beta ,\mu} \frac{d x^\alpha}{d \lambda} \frac{d x^\beta}{d \sigma}+g_e K A_{\alpha,\mu} \frac{d x^\alpha}{d \lambda}
\end{equation}

\noindent after using Eq.\eqref{mande1}, so $L^{GUGE}$ gives rise to the usual equations of motion \eqref{eqmchg} for a {\it {charged}} particle of (dynamically conserved) charge $K$ (which may be either positive or negative), instead of the external parameter $q$. Note that the derivatives are taken with respect to $\sigma$.

The 5--dimensional GUGE Lagrangian $L^{GUGE}$ produces (geodesic, in 5--dimensions) equations of motion which are completely equivalent to the usual (non--geodesic) ones for a charged particle in 4--dimensional spacetime \eqref{eqmchg}.

\begin{table*}[htbp]
  \centering
  \begin{tabular}{ |p{2.0cm}||p{8.5cm}|p{7.0cm}|  }
    \hline \hline
     &   & \\
    & KK & GUGE \\
    &   & \\
    \hline \hline
    Assumption 1  &  KK metric & Proper time redefinition Lagrangian \\
     \hline
    Assumption 2   & Cylinder condition & {\bf {-- -- -- -- -- -- --}}\\
     \hline
     Assumption 3   & Compactification condition & {\bf {-- -- -- -- -- -- -- }}\\
    \hline
    &   & \\
     \hline
    Outcome 1  &{\bf {-- -- -- -- -- -- --}} &  GUGE metric\\
     \hline
     Outcome 2  & {\bf {-- -- -- -- -- -- --}} &  Cylinder condition\\
    \hline
    Outcome 3  & {\bf {-- -- -- -- -- -- -- }} &  Compactification condition\\
     \hline
     Outcome 4  & Field equations similar to those of Einstein--Maxwell &  Einstein--Maxwell field equations\\
     \hline
     Outcome 5  & Equations similar to those of relativistic charged particles & Relativistic charged particle equations \\
     \hline
     Outcome 6  & {\bf {-- -- -- -- -- -- -- }} & Electric charge conservation\\
     \hline
     Outcome 7  & Mass tower & Electric charge quantization\\
        \hline
     Outcome 8 & {\bf {-- -- -- -- -- -- -- }} & Electric charge contribution to particle's energy\\
     \hline
     Outcome 9 & {\bf {-- -- -- -- -- -- -- }} & There are no massless charged particles\\

        \hline  \hline
  \end{tabular}
  \caption{Comparison between KK and GUGE}
  \label{tab: 1}
\end{table*}

\section{Further Results}
Consider the simplest flat spacetime and no electromagnetic field case. The (5--)momentum constraint equation (reintroducing $m_0$, for the sake of clarity) reads
\begin{equation}
({g^{G}})^{ab} P_a P_b \equiv - {P_0}^2+{P_1}^2+{P_2}^2+{P_3}^2+{P_\phi}^2=-{m_0}^2,  \label{5mom1}
\end{equation}

or

\begin{equation}
 P_0 = \pm \left({P_1}^2+{P_2}^2+{P_3}^2+K^2+{m_0}^2\right)^{1/2},  \label{5mom2}
\end{equation}
which clearly shows that, even in the absence of electromagnetic fields, the electric charge contributes to the energy of a charged particle. Therefore, the rest energy of the particle (for $P_1=P_2=P_3=0$) does not vanish if $K \neq 0$ even if $m_0$ vanishes, which means that there are no massless electrically charged particles, as observed.\\

This result may be interpreted as a redefinition of the rest mass for charged particles.\\

The Klein--Gordon equation associated to \eqref{5mom1}, the fact that $0\leq \phi \leq 2 \pi$ plus the usual requirement that the wave function be single valued leads to (its canonically conjugated momentum $K$) charge quantization, namely, $K = n\ (= 0, \pm 1, \pm 2, ...$ ), which is, of course, observed. This quantization happens much in the same way as the $z$--direction angular momentum quantization for rotationally symmetric potentials in quantum mechanics. This same argument has led in the prevailing treatment of Kaluza--Klein theory to the so called ``mass towers'', using {\it {ad hoc}} cylinder and compactification conditions for an artificially defined cyclical length--like $5^{th}$ dimensional coordinate \cite{grard}.\\

\section{Summary and Outlook}

A GUGE based on a Riemann metric $g^F$  is presented. The Riemann metric is constructed from a newly defined Lagrangian \cite{propertime} which describes the dynamics of a point massive test particle interacting with gravitation, electromagnetism and Yang--Mills fields. The new Lagrangian, which is similar to the usual one, is created by requiring that it be gauge invariant under the usual electromagnetic and Yang--Mills gauge transformations (unlike the usual one). This gauge invariance is achieved by introducing new internal complex coordinates which are associated to electromagnetic and Yang--Mills charges. This requirement is based on the condition that its action be proportional to proper time even when the non--gravitational fundamental interactions are present \cite{propertime}. Its equations of motion include the usual ones and, in addition, predict electric charge conservation. The 5--dimensional GUGE metric for the case of unification of gravitational and electromagnetic interactions may be written as $g^F=g^{G}$  where $g^{G}$ {\it {looks}} identical to the usual 5--dimensional Kaluza--Klein metric but its different in several respects. To get the GUGE theory, one may construct the Einstein--Hilbert Lagrangian based on $g^G$.

One may consider this approach as a way of getting GUGE theory metric instead of postulating it as it is done in the Kaluza--Klein approach. The GUGE theory is completely equivalent to Maxwell--Einstein theory both for the (geodesic) motion of point particles in the GUGE background metric and as a theory to describe interacting gravitational and electromagnetic fields.

In addition, there is no need to invoke either the cylinder or the curling of coordinates conditions, as they are consequences of the GUGE theory. Furthermore, no extra scalar field appears.  Note that the coordinate $Q$ disappears completely in the unified description of gravitation and electromagnetism. It is important to emphasize that the fifth dimension has a clear physical interpretation in this approach.

The Riemann metric in which is based, yields a redefinition of proper time whose possible implications and experimental verification are presented in \cite{propertime}. Lorentz transformations in addition to translations may also be incorporated to the unification of fundamental interactions using this scheme (see, \cite{propertime}).

A comparison between GUGE and KK seems in order. one important point is that the GUGE metric is deduced from results obtained in \cite{propertime} while the KK metric is postulated. In the GUGE field theory presented here, the ``cylinder'' and the ``curling of coordinates'' emerge as direct consequences of the construction of the GUGE metric instead of being postulated as it is done in the KK theory. The GUGE field equations are completely equivalent to the Einstein--Maxwell field equations, while KK field equations are not. The GUGE 5--dimensional (Riemannian geodesic) equations are fully equivalent to the 4--dimensional (non--geodesic) equations for a charged particle moving in the presence of gravitational and electromagnetic fields, unlike the KK 5--dimensional (geodesic) equations which are not and, in addition, yield non--gauge covariant geodesic equations. The Kaluza--Klein theory has an extra scalar field with no clear physical meaning while no extra scalar field appears in GUGE. Finally, the physical interpretation of the fifth dimension and of the role of the extra field in KK (internal coordinate in GUGE) are totally different in both approaches.\\

In addition to electric charge conservation, the GUGE approach yields electric charge quantization and electric charge contribution to the particle's energy which means the {\it{there are no massless electrically charged particles in Nature}}, as it has been observed, so far.\\

Future research includes the study of quantum aspects of the GUGE proposal, computing the magnitude (and the feasibility of observation) of the predictions associated to the new definition of proper time, to the electric charge dependence of the particle's energy and to the unification scheme presented here. The inclusion of Yang--Mills interactions is also the subject of ongoing work.\\

\begin{acknowledgments}
It is a pleasure to thank Felipe A. Asenjo and Benjamin Koch for very many enlightening discussions. The author also wishes to express his gratitude to Matthew Dawson, Andr\'es Gomberoff, Francisco Rojas, Oscar Adolfo S\'anchez--Valenzuela and Gianni Tallarita.
\end{acknowledgments}

\end{document}